\documentclass{article}
\usepackage{amsfonts}
\usepackage{amsmath}

\input{tcilatex}

\begin{document}

\title{Bare and dressed particles in collision theory}
\author{Giovanni Falcone \\
Universit\`{a} della Calabria, Dip. di Fisica, 87036 Rende (Cs) Italy\\
and Gruppo collegato INFN di Cosenza}
\maketitle

\begin{abstract}
The bare-dressed technique is, for the first time, used in collision theory.
The approach is valid for classical as well for quantum binary elastic
collisions in the non relativistic regime. The same formalism can be used
for inelastic collisions, when the particles undergo only a change of their
internal quantum states during the collision process. All kinematic results
can be obtained by a simple matrix transformation. Moreover, to make also
simple and clear the results, for the inelastic collisions, the restitution
coefficient formulation of inelastic processes has been used.



\textbf{e-mail address}: falcone@fis.unical.it
\end{abstract}

\section{Introduction}

The idea of transforming a system of interacting real particles into a
system of non-interacting fictitious particles has been used in different
fields of the physics. Concepts like bare and dressed particles, bare and
dressed states, bare and dressed interactions and so on, can be found in
many formulations of physical problems.\cite{Mat, Mey} The basic idea of the
bare-dressed approach is the following. Once the formulation of the physical
problem for the real objects has been obtained one has to look for a
transformation that enables to reformulate the original physical problem to
that one for fictitious (dressed) objects. The modified formulations are
more simple, in general. Once the more simple solutions for the dressed
objects have been obtained, through an inverse transformation one get the
solutions for the real objects. Bare and dressed physical objects can be
found in the classical as well in quantum physics,\cite{Mat} although it is,
in quantum physics, that the bare-dressed techniques have received the more
interesting applications. In this paper, we discuss of bare and dressed
particles in collision theory, in the non relativistic regime. Moreover, it
is well known \cite{Joa} that the obtained results are valid for classical
as well for quantum physics.

One of the most important kinematical problems, in collision theory,
consists in transforming the results of measurements and calculations from
one frame of reference to another.\cite{Joa,Fal1} The laboratory and the
center of mass systems are the most used frames of reference. Although the
two frames of references are important, the bare-dressed approach underlines
the central role of the transformation from the real to the dressed
particles in collision theory. Moreover, we shall show that the bare-dressed
technique is so natural for binary collisions that all kinematic results can
be obtained in a very simple and general way.

We first introduce, through the two body problem, the concept of bare and
the dressed particles, and subsequently we apply the bare-dressed technique
to the elastic and inelastic collisions in the non relativistic regime.

\section{Bare and dressed particles in the two body problem}

Quantum mechanically, as classically, a very important class of dynamical
problems arises with the discussion of an isolated two-particle system. For
this system the Lagrangian can be expressed as%
\begin{equation}
L=\frac{M_{1}}{2}\left\vert \overrightarrow{v}_{1}\right\vert ^{2}+\frac{%
M_{2}}{2}\left\vert \overrightarrow{v}_{2}\right\vert ^{2}-U\left( 
\overrightarrow{r}_{1}-\overrightarrow{r}_{2}\right)  \label{1}
\end{equation}%
where $\left( \overrightarrow{r}_{1},\overrightarrow{v}_{1}\right) $ and $%
\left( \overrightarrow{r}_{1},\overrightarrow{v}_{1}\right) $ represent the
position and the velocity of the particles of mass $M_{1}$ and $M_{2}$,
respectively and $U\left( \overrightarrow{r}_{1}-\overrightarrow{r}%
_{2}\right) $ is the potential energy for the isolated two-particle system.

In order to solve the previous problem the position $\overrightarrow{R}$ of
the center of mass is introduced%
\begin{equation}
\overrightarrow{R}=\frac{M_{1}\overrightarrow{r}_{1}+M_{2}\overrightarrow{r}%
_{2}}{M_{1}+M_{2}}  \label{2}
\end{equation}%
together with the inter-particle vector 
\begin{equation}
\overrightarrow{r}=\overrightarrow{r}_{1}-\overrightarrow{r}_{2}  \label{3}
\end{equation}%
As a consequence of the previous transformations, the Lagrangian of the
isolated two-particle system becomes\cite{Fal1} 
\begin{equation}
L=\frac{M}{2}\left\vert \overrightarrow{V}\right\vert ^{2}+\frac{\mu }{2}%
\left\vert \overrightarrow{v}\right\vert ^{2}-U\left( \overrightarrow{r}%
\right)  \label{4}
\end{equation}

where 
\begin{equation}
M=M_{1}+M_{2}\qquad \mu =\frac{M_{1}M_{2}}{M_{1}+M_{2}}  \label{5}
\end{equation}%
are the total and the reduced mass of the two particles, respectively.
Moreover, $\overrightarrow{V}=d\overrightarrow{R}/dt$ and $\overrightarrow{v}%
=d\overrightarrow{r}/dt$. \ Since for an isolated system the canonical
momentum%
\begin{equation}
\overrightarrow{P}=\frac{\partial L}{\partial \overrightarrow{V}}  \label{6}
\end{equation}%
is a constant of the motion, in the center of mass system of reference, the
Lagrangian is reduced to%
\begin{equation}
L=\frac{\mu }{2}\left\vert \overrightarrow{v}\right\vert ^{2}-U\left( 
\overrightarrow{r}\right)  \label{7}
\end{equation}

The two-body problem has been reduced to one-body problem. A similar result
can be obtained by using the hamiltonian or the newtonian formulation. All
formulations share the two transformations, eqs. (\ref{2})and (\ref{3}).
Therefore eqs. (\ref{2})and (\ref{3}) represent the (true) basic quantities
of the previous results.These transformations enable to obtain two
fictitious particles $M$ and $\mu $ from the two real particles $M_{1}$ and $%
M_{2}$.

Our approach to collision theory is based on eqs. (\ref{2})and (\ref{3}),
and we recognize as \textit{bare}, the real particles and as \textit{dressed}%
, the fictitious particles, $M$ and $\mu $.

Since transformations are better represented by a matrix formulation we can
write eqs. (\ref{2})and (\ref{3}) together as

\begin{equation}
\left( 
\begin{array}{c}
\overrightarrow{R} \\ 
\overrightarrow{r}%
\end{array}%
\right) =U\left( 
\begin{array}{c}
\overrightarrow{r}_{1} \\ 
\overrightarrow{r}_{2}%
\end{array}%
\right)  \label{8}
\end{equation}%
where the matrix $U$ is%
\begin{equation}
U=\left( 
\begin{array}{cc}
\frac{M_{1}}{M_{1}+M_{2}} & \frac{M_{2}}{M_{1}+M_{2}} \\ 
1 & -1%
\end{array}%
\right)  \label{9}
\end{equation}%
It is not difficult to prove that the inverse matrix $U^{-1}$ is%
\begin{equation}
U^{-1}=\left( 
\begin{array}{cc}
1 & \frac{M_{2}}{M_{1}+M_{2}} \\ 
1 & -\frac{M_{1}}{M_{1}+M_{2}}%
\end{array}%
\right)  \label{10}
\end{equation}%
In the next section we present a matrix formulation of the binary kinematic
collisions in the non relativistic regime based on the matrix $U$ and its
inverse.

\section{The matrix formulation of the binary kinematic collisions}

In the discussion of the binary collisions, the kinematic aspects are
separated from the collision dynamic, where the particular nature of the
interaction is taken into account. In fact, collision kinematics are
described in terms of conservation laws without any specification of the
nature of the interaction. In particular, for the elastic collisions one
writes the energy and momentum conservation laws:%
\begin{equation}
M_{1}\overrightarrow{v}_{1}+M_{2}\overrightarrow{v}_{2}=M_{1}\overrightarrow{%
v}_{1}^{\prime }+M_{2}\overrightarrow{v}_{2}^{\prime }  \label{11}
\end{equation}%
\begin{equation}
\frac{1}{2}M_{1}v_{1}^{2}+\frac{1}{2}M_{2}v_{2}^{2}=\frac{1}{2}%
M_{1}v_{1}^{\prime 2}+\frac{1}{2}M_{2}v_{2}^{\prime 2}  \label{12}
\end{equation}%
where the prime indicate the quantities after the collision. The problem to
be solved is to find the velocities of the particles after the collision in
terms of those ones before the collision. The general solution is not easy
to find in a simple way and the most part of the books prefer to present the
solution for the laboratory system of reference, where the previous
equations become:%
\begin{equation}
M_{1}\overrightarrow{v}_{1}=M_{1}\overrightarrow{v}_{1}^{\prime }+M_{2}%
\overrightarrow{v}_{2}^{\prime }  \label{13}
\end{equation}%
\begin{equation}
\frac{1}{2}M_{1}v_{1}^{2}=\frac{1}{2}M_{1}v_{1}^{\prime 2}+\frac{1}{2}%
M_{2}v_{2}^{\prime 2}  \label{14}
\end{equation}

We shall derive the solution of the general problem by using $U$ and its
inverse.

In the non relativistic regime, velocities involved in collisions can be
obtained by a simple derivative with respect to time of the position
vectors. Therefore, we can also write%
\begin{equation}
\left( 
\begin{array}{c}
\overrightarrow{V} \\ 
\overrightarrow{v}%
\end{array}%
\right) =U\left( 
\begin{array}{c}
\overrightarrow{v}_{1} \\ 
\overrightarrow{v}_{2}%
\end{array}%
\right)  \label{15}
\end{equation}%
namely, the matrix $U$ transforms also the velocities of the bare particles,
before the collision, into the velocities of the dressed particles, before
the collision. This is the great advantage of the non relativistic regime.
It is not difficult to prove that the inverse matrix $U^{-1}$ will transform
the velocities of the dressed particles, after the collision, into the bare
particles, after the collision:%
\begin{equation}
\left( 
\begin{array}{c}
\overrightarrow{v}_{1}^{\prime } \\ 
\overrightarrow{v}_{2}^{\prime }%
\end{array}%
\right) =U^{-1}\left( 
\begin{array}{c}
\overrightarrow{V}^{\prime } \\ 
\overrightarrow{v}^{\prime }%
\end{array}%
\right)  \label{16}
\end{equation}

Eqs.(\ref{9}) and (\ref{10}) are valid for all binary elastic collisions, in
the non relativistic regime, and kinematic results depend only on the
determination of $\overrightarrow{V}^{\prime }$ and $\overrightarrow{v}%
^{\prime }$.

Since during the collision the $M$ particle cannot change the velocity, we
can write%
\begin{equation}
\overrightarrow{V}^{\prime }=\overrightarrow{V}  \label{17}
\end{equation}

This result is valid for both elastic and inelastic collisions. Therefore,
the difference between the two regimes depends only on $\overrightarrow{v}%
^{\prime }$.

In the case of \textit{elastic} collisions the kinetic energy is conserved:%
\begin{equation}
\frac{1}{2}\mu v^{\prime 2}=\frac{1}{2}\mu v^{2}  \label{18}
\end{equation}

Moreover, by definition of collisions $\overrightarrow{v}^{\prime }\neq 
\overrightarrow{v}$: the direction of the velocity $\overrightarrow{v}$ $\ $%
is changed or it is changed its sign during the collision process. Then, one
obtains%
\begin{equation}
\overrightarrow{v}^{\prime }=-\overrightarrow{v}  \label{19}
\end{equation}

Since we have the velocities of the dressed particles after the collision,
we can use eq.(\ref{10}) and, by a simple transformation write%
\begin{equation}
\overrightarrow{v}_{1}^{\prime }=\frac{M_{1}-M_{2}}{M_{1}+M_{2}}%
\overrightarrow{v}_{1}+\frac{2M_{2}}{M_{1}+M_{2}}\overrightarrow{v}_{2}
\label{20}
\end{equation}%
\begin{equation}
\overrightarrow{v}_{2}^{\prime }=\frac{2M_{1}}{M_{1}+M_{2}}\overrightarrow{v}%
_{1}-\frac{M_{1}-M_{2}}{M_{1}+M_{2}}\overrightarrow{v}_{2}  \label{21}
\end{equation}

These are the general solutions for the elastic regime.

If we multiply eq.(\ref{20}) for $M_{1}$ and eq.(\ref{21}) for $M_{2}$ we
obtain

\begin{equation}
\overrightarrow{p}_{1}^{\prime }=-\mu \overrightarrow{v}+\frac{M_{2}}{%
M_{1}+M_{2}}\left( \overrightarrow{p}_{1}+\overrightarrow{p}_{2}\right)
\label{22}
\end{equation}

\begin{equation}
\overrightarrow{p}_{2}^{\prime }=\mu \overrightarrow{v}+\frac{M_{1}}{%
M_{1}+M_{2}}\left( \overrightarrow{p}_{1}+\overrightarrow{p}_{2}\right)
\label{23}
\end{equation}

In this form the kinematic results are often presented and discussed.\cite%
{Fal1} For the \textit{laboratory system of reference} ($\overrightarrow{v}%
_{2}=0$) we obtain%
\begin{equation}
\overrightarrow{p}_{1}^{\prime }=-\mu \overrightarrow{v}_{1}+\frac{M_{2}}{%
M_{1}+M_{2}}\overrightarrow{p}_{1}  \label{24}
\end{equation}%
\begin{equation}
\overrightarrow{p}_{2}^{\prime }=\mu \overrightarrow{v}_{1}+\frac{M_{1}}{%
M_{1}+M_{2}}\overrightarrow{p}_{1}  \label{25}
\end{equation}

whereas for the \textit{center of mass system of reference} (the total
momentum of the system is zero) we get%
\begin{equation}
\overrightarrow{\widetilde{p}}_{1}^{\prime }=-\mu \overrightarrow{v}
\label{26}
\end{equation}%
\begin{equation}
\overrightarrow{\widetilde{p}}_{2}^{\prime }=\mu \overrightarrow{v}
\label{27}
\end{equation}

where we have used the \textit{tilde} for the quantities measured in the
center of mass system of reference. In conclusion, all kinematic results in
the elastic regime can be reproduced by using the inverse matrix together
with eqs. (\ref{17}) and (\ref{19}).

\section{The one dimensional inelastic collisions}

The previous approach can be extended to the inelastic collisions under the
assumption that the particles undergo only a change of their internal
quantum states during the collision process. \ Since the mass of the two
particles can be considered constant we can use again eqs. (\ref{9}) and (%
\ref{10}). Nevertheless, since the inelastic loss may be anisotropic we
consider, in this paper, only the one dimensional case. Moreover, to make a
comparison with the elastic collisions more direct, let us consider a
definite one dimensional collision. Let us assume that the particles, before
the collision, move along the x axis. We can write, first,%
\begin{equation}
V_{x}=\frac{M_{1}v_{x1}+M_{2}v_{x2}}{M_{1}+M_{2}}\qquad v_{x}=v_{x1}-v_{x2}
\label{28}
\end{equation}%
then%
\begin{equation}
\left( 
\begin{array}{c}
v_{x1}^{\prime } \\ 
v_{x2}^{\prime }%
\end{array}%
\right) =U^{-1}\left( 
\begin{array}{c}
V_{x} \\ 
-v_{x}%
\end{array}%
\right)  \label{29}
\end{equation}%
and finally%
\begin{equation}
v_{x1}^{\prime }=\frac{M_{1}-M_{2}}{M_{1}+M_{2}}v_{x1}+\frac{2M_{2}}{%
M_{1}+M_{2}}v_{x2}  \label{30}
\end{equation}%
\begin{equation}
v_{x2}^{\prime }=\frac{2M_{1}}{M_{1}+M_{2}}v_{x1}-\frac{M_{1}-M_{2}}{%
M_{1}+M_{2}}v_{x2}  \label{31}
\end{equation}%
These results can be obtained directly from the general solutions, eqs. (\ref%
{20}) and (\ref{21}).

Let us consider the equivalent inelastic collision. We shall adopt, in the
description of the inelastic collisions, the restitution coefficient
approach ,\cite{Fal2,Fal3} because this approach, as we shall show, makes
simple the results and their comparison with the elastic results.

The velocity of $M$ particle is still unchanged, whereas the energy
conservation can be written,\cite{Fal2,Fal3} as follows 
\begin{equation}
\frac{1}{2}\mu v_{x}^{\prime 2}=\chi ^{2}\frac{1}{2}\mu v_{x}^{2}  \label{32}
\end{equation}%
The conservation of the momentum tell us that 
\begin{equation}
v_{x}^{\prime }=-\chi v_{x}  \label{33}
\end{equation}

Since the $U$ matrix and its inverse are not changed, we arrive immediately
at following results 
\begin{equation}
v_{x1}^{\prime }=\frac{M_{1}-\chi M_{2}}{M_{1}+M_{2}}v_{x1}+\frac{\left(
1+\chi \right) M_{2}}{M_{1}+M_{2}}v_{x2}  \label{34}
\end{equation}%
\begin{equation}
v_{x2}^{\prime }=\frac{\left( 1+\chi \right) M_{1}}{M_{1}+M_{2}}v_{x1}+\frac{%
M_{2}-\chi M_{1}}{M_{1}+M_{2}}v_{x2}  \label{35}
\end{equation}%
In the case of elastic collisions, $\chi =1$, and we find again eqs. (\ref%
{20}) and (\ref{21}).

\section{Conclusions}

The use of the bare-dressed technique has been exploited, for the first
time, in the description of binary collisions. In this approach the basic
quantities are the transformations from the real to the dressed particles.
These transformations are valid for all binary collisions when the particles
do not change their masses during the collision process. Moreover, in the
case of inelastic collisions, the base-dressed formulation together with the
restitution coefficient approach make the inelastic results a natural
extension of the elastic ones. The bare-dressed technique is of general
character and enables to obtain all binary kinematic results in the non
relativistic regime for classical as well for quantum physics.

\end{document}